\documentclass[conference]{IEEEtran}
\IEEEoverridecommandlockouts
\usepackage{amsmath,amssymb,amsfonts}
\usepackage{algorithmic}
\usepackage{graphicx}
\usepackage{textcomp}
\usepackage{xcolor}
\usepackage{tikz}
\usepackage{float}
\usepackage{booktabs}
\usepackage{multirow}
\usepackage{array}
\usepackage{xcolor}

\usepackage[T1]{fontenc}
\usepackage[utf8]{inputenc}
\usepackage{listings}

\usepackage[
backend=biber,
style=alphabetic,
sorting=ynt
]{biblatex}

\usepackage{hyperref}  % Add this package for clickable links
\hypersetup{
	colorlinks=true,    % Colors the text of links
	linkcolor=blue,     % Color of internal links
	urlcolor=cyan,      % Color of external links
	citecolor=green     % Color of citations
}
\usetikzlibrary{shapes.geometric, arrows}

\definecolor{keywordcolor}{rgb}{0.7, 0.1, 0.1}   % red
\definecolor{tacticcolor}{rgb}{0.0, 0.1, 0.6}    % blue
\definecolor{commentcolor}{rgb}{0.4, 0.4, 0.4}   % grey
\definecolor{symbolcolor}{rgb}{0.0, 0.1, 0.6}    % blue
\definecolor{sortcolor}{rgb}{0.1, 0.5, 0.1}      % green
\definecolor{attributecolor}{rgb}{0.7, 0.1, 0.1} % red

\DeclareUnicodeCharacter{03BB}{\ensuremath{\lambda}}
\DeclareUnicodeCharacter{2265}{\ensuremath{\geq}}

\begin{document}

\title{Proving the Coding Interview: A Benchmark for Formally Verified Code Generation}%\thanks{Funded by Anthropic research credits.}}
% I moved thanks back up here because the original template had it here. 

\author{\IEEEauthorblockN{Quinn Dougherty}
\IEEEauthorblockA{\textit{Beneficial AI Foundation} \\
quinn.dougherty92@gmail.com}
\and
\IEEEauthorblockN{Ronak Mehta}
\IEEEauthorblockA{\textit{Independent} \\
ronakrm@gmail.com}
}

\maketitle

\begin{abstract}
We introduce the Formally Verified Automated Programming Progress Standards, or FVAPPS, a benchmark of 4715 samples for writing programs and proving their correctness, the largest formal verification benchmark, including 1083 curated and quality controlled samples. Previously, APPS provided a benchmark and dataset for programming puzzles to be completed in Python and checked against unit tests, of the kind seen in technical assessments in the software engineering industry. Building upon recent approaches for benchmarks in interactive theorem proving, we generalize the unit tests to Lean 4 theorems given without proof (i.e., using Lean's ``sorry'' keyword). On the 406 theorems of 100 randomly selected samples, Sonnet correctly proves 30\% and Gemini correctly proves 18\%. We challenge the machine learning and program synthesis communities to solve both each general purpose programming problem and its associated correctness specifications. The benchmark is available at \url{https://huggingface.co/datasets/quinn-dougherty/fvapps}. 
\end{abstract}


\section{Introduction}

Large Language Models (LLMs) have developed incredible capabilities, spanning creative writing, tutoring and education, information retrieval and distillation, common sense reasoning, complex mathematical reasoning, and even code generation \cite{chatbotarena,gpt4,llama3,deepseek}. Particular focus and attention has been given towards LLM-assisted software development, given its economic value and position between natural language and formal logic.

We have seen rapid advancements in code generation capabilities~\cite{Singh2023CodeFusionAP,jimenez2023swe,Hendrycks2021MeasuringCC}. With this has come strong concerns around the misuse and safety of code generated by these models, and the ability to prove or guarantee that the produced code is sound. An approach commonly employed today includes running LLM-generated code in restricted development environments, and ensuring tight feedback loops with software engineers who understand the unique failure modes of their particular complex software. This paired programming-style approach is common in LLM chatbot interfaces, as well as programming assistants more strongly integrated in the software engineering development cycle such as GitHub Copilot~\cite{chen2021evaluating} and Cursor (Anysphere Inc., 2023).

However, recent paradigm shifts in the ability to develop and enhance capabilities through supervised fine-tuning has led to fully autonomous software engineering, which includes code generation \textit{and execution} in the development loop (e.g., Devin, Cognition AI, Inc. 2023, \cite{wang2024openhandsopenplatformai}).  Without a human evaluating code before it is run, strong assumptions or restrictions are necessary to ensure that generated code is safe and correct. 
Critically, as these tools become ubiquitous across more and more applications that touch the real world, it will be a requirement that we have guarantees of their safety and correctness.
Code generated for managing medical devices, for managing patient records, for ensuring the security of private personal information, for controlling autonomous vehicles, for cybersecurity defense systems, and for military use will all require strong guarantees before users, organizations, and governments can make effective use of these tools.

While software quality assurance primarily relies on extensive testing, the emerging field of deductive verification offers a more rigorous alternative by mathematically proving code correctness against formal specifications.
However, this approach requires deep system knowledge, explicit property definitions, and often demands rewriting existing code in specialized proof assistant languages—a significant challenge given the current state of tools.
As Large Language Models (LLMs) advance in code generation capabilities, they may eventually enable automated creation of deductively verifiable code, allowing us to defer trust to theorem provers rather than testing alone. This could revolutionize mission-critical software development by providing formal guarantees against bugs and hallucinations, though current LLM capabilities fall short of this goal, and it remains unclear whether fine-tuning or prompt engineering can bridge this gap.

\noindent\textbf{Contributions.} We deliver a set of 4,715 Lean 4 files consisting of a solution function signature and theorem statements, all of which are passed on for future proving via the ``sorry'' keyword, allowing the \texttt{lean} executable to compile without actual proofs or implementations. We provide tags for samples that pass additional, rigorous quality assurance, defined by inline unit testing and property tests in Lean.

\section{The FVAPPS Benchmark}

\begin{figure*}
    
    \centering
    \includegraphics[width=\textwidth]{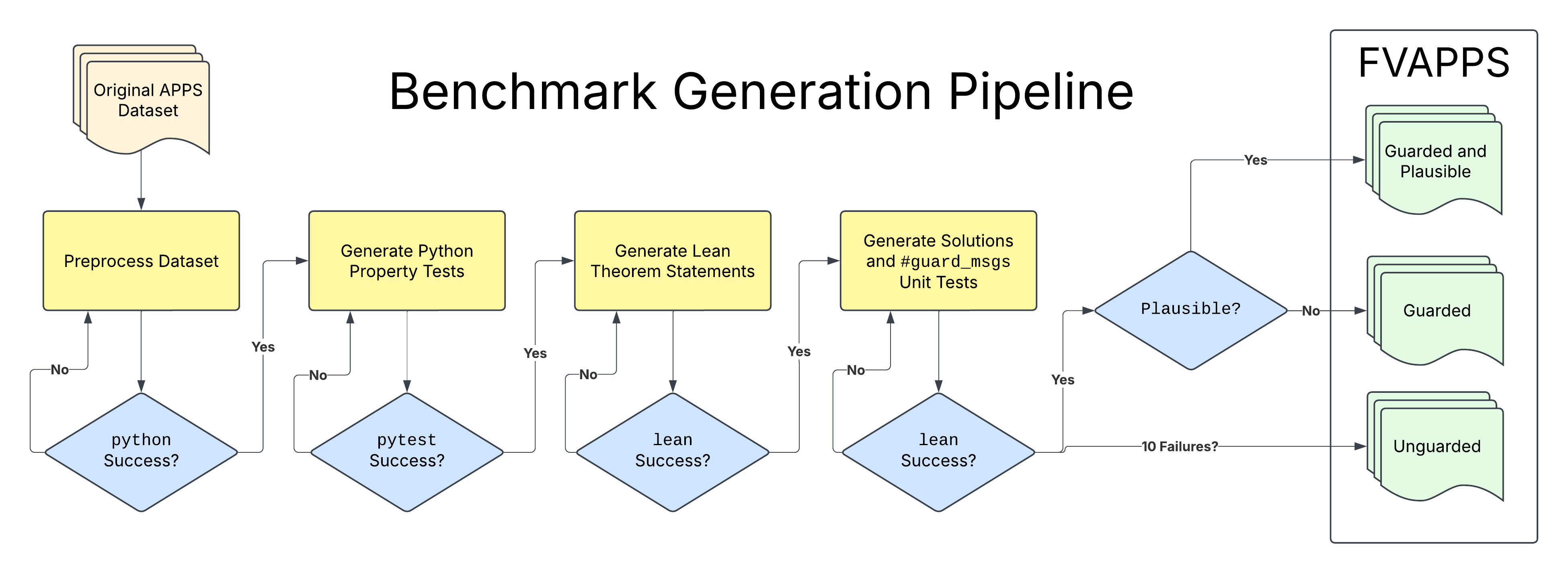}
    \caption{Benchmark generation pipeline for creating coding interview theorem statements in Lean from APPS questions and solutions.}
    \label{fig:benchmark-pipeline}
\end{figure*}

\newcommand{\pytestloop}{
    \begin{tikzpicture}[
        auto,
        block/.style = {draw, circle, minimum size=1cm},
        smallblock/.style = {draw, rectangle, minimum size=0.7cm},
        line/.style = {draw, -stealth, thick}
    ]

    % Nodes
    \node [block] (api) at (0,0) {Sonnet 3.5};
    \node [block] (pytest) at (2,-2) {$\texttt{pytest}$};
    \node [block] (lc) at (-1,-4) {$\lambda c.c+\texttt{stderr}$};
    \node [smallblock] (exit) at (2,-4) {exitcode};
    \node [block] (write) at (4,-5) {$\textit{write}$};

    % Connections
    \draw [line] (-3,0) -- node [above] {$p_0$} (api);
    \path [line] (api) to [bend left] node [above] {$c_k$} (pytest);
    \path [line] (pytest) -- node [above] {} (exit);
    \path [line] (exit) -- node [above] {0} (write);
    \path [line] (exit) -- node [sloped, above, midway] {1} (lc);
    \path [line] (lc) to [bend left] node [left] {$p_{k+1}$} (api);

    \end{tikzpicture}
}

\newcommand{\leanloop}{
    \begin{tikzpicture}[
        auto,
        block/.style = {draw, circle, minimum size=1cm},
        smallblock/.style = {draw, rectangle, minimum size=0.7cm},
        line/.style = {draw, -stealth, thick}
    ]

    % Nodes
    \node [block] (api) at (2,0) {Sonnet 3.5};
    \node [block] (read) at (-1,0) {$\textit{read}$};
    \node [block] (lean) at (2,-2) {$\texttt{lean}$};
    \node [block] (lc) at (-1,-4) {$\lambda c.c+\texttt{stderr}$};
    \node [smallblock] (exit) at (2,-4) {exitcode};
    \node [block] (done) at (4,-5) {$\textit{write} (done)$};

    % Connections
    \draw [line] (-3,0) -- node [above] {} (read);
   % \path [line] (read) -- (api) [above] {$p_0$};
    \path [line] (read) to node [above] {$p_0$} (api);
    \path [line] (api) to [bend left] node [right] {$c_k$} (lean);
    \path [line] (lean) -- node [above] {} (exit);
    \path [line] (exit) -- node [above] {0} (done);
    \path [line] (exit) -- node [sloped, above, midway] {1} (lc);
    \path [line] (lc) to [bend left] node [left] {$p_{k+1}$} (api);

    \end{tikzpicture}
}

\newcommand{\genericloop}{
    \begin{tikzpicture}[
        auto,
        block/.style = {draw, circle, minimum size=1cm},
        smallblock/.style = {draw, rectangle, minimum size=0.7cm},
        line/.style = {draw, -stealth, thick}
    ]

    % Nodes
    \node [block] (api) at (2,0) {Sonnet 3.5};
    \node [block] (read) at (-1,0) {$\textit{read}$};
    \node [block] (run) at (2,-2) {$\texttt{run}$};
    \node [block] (lc) at (-1,-4) {$\lambda c.c+\texttt{stderr}$};
    \node [smallblock] (exit) at (2,-4) {exitcode};
    \node [block] (done) at (4,-5) {$\textit{write}$};

    % Connections
    \draw [line] (-3,0) -- node [above] {} (read);
   % \path [line] (read) -- (api) [above] {$p_0$};
    \path [line] (read) to node [above] {$p_0$} (api);
    \path [line] (api) to [bend left] node [right] {$c_k$} (run);
    \path [line] (run) -- node [above] {} (exit);
    \path [line] (exit) -- node [above] {0} (done);
    \path [line] (exit) -- node [sloped, above, midway] {1} (lc);
    \path [line] (lc) to [bend left] node [left] {$p_{k+1}$} (api);

    \end{tikzpicture}
}

A strong benchmark for evaluating model capabilities for theorem proving should have a number of desirable properties. With a particular focus on software engineering, algorithms and properties of those algorithms should well represent tasks and solutions that are used to evaluate skill in the labor force today. Coding interview questions fit this requirement naturally.
With generic coding capabilities well-evaluated,
we can build upon existing benchmarks of coding ability
to generate our formally-verified one.

We begin with APPS, a benchmark and dataset for programming puzzle solving \cite{Hendrycks2021MeasuringCC}. Developed largely by scraping the interview preparation platforms CodeForces and LeetCode, each sample consists of a natural language question, a ternary flag marking difficulty, a list of correct solutions in Python, and a few input-output test case pairs. 

\subsection{An Example Task}

\begin{figure*}
    \centering
\begin{lstlisting}
def solve_elections (n : Nat) (voters : List (Nat × Nat)) : Nat := sorry

theorem solve_elections_nonnegative (n : Nat) (voters : List (Nat × Nat)) : solve_elections n voters >= 0 := sorry

theorem solve_elections_upper_bound (n : Nat) (voters : List (Nat × Nat)) : solve_elections n voters <= List.foldl (λ acc (pair : Nat × Nat) => acc + pair.2) 0 voters := sorry

theorem solve_elections_zero_votes (n : Nat) (voters : List (Nat × Nat)) : (List.all voters (fun pair => pair.1 = 0)) -> solve_elections n voters = 0 := sorry

theorem solve_elections_single_zero_vote : solve_elections 1 [(0, 5)] = 0 := sorry
\end{lstlisting}
    \caption{FVAPPS sample 23, derived from train sample 23 of APPS source. The \texttt{def} is where the solver implements the function, each \texttt{theorem} is a correctness specification.}
    \label{fig:sample0023}
\end{figure*}

As an example, consider the crooked election problem:
\begin{quote}
There are $n$ voters, and two ways to convince each of them to vote for you. The first way to convince the $i$-th voter is to pay him $p_i$ coins. The second way is to make $m_i$ other voters vote for you, and the $i$-th voter will vote for free.
Moreover, the process of such voting takes place in several steps. For example, if there are five voters with $m_1 = 1$, $m_2 = 2$, $m_3 = 2$, $m_4 = 4$, $m_5 = 5$, then you can buy the vote of the fifth voter, and eventually everyone will vote for you. Set of people voting for you will change as follows: ${5} \rightarrow {1, 5} \rightarrow {1, 2, 3, 5} \rightarrow {1, 2, 3, 4, 5}$.

Calculate the minimum number of coins you have to spend so that everyone votes for you.
\end{quote}
To get everyone to vote for you, you can either purchase votes for each voter's declared price, or convert each voter through their declared groupthink threshold. This is the setup of sample \texttt{0023}, derived from APPS problem train/23\footnote{https://huggingface.co/datasets/codeparrot/apps/viewer/all/train?row=23}. Figure \ref{fig:sample0023} is our pipeline's output that we ship in our dataset.
Crucially, it is one \texttt{def} and multiple \texttt{theorem}s, though many samples include extra definitions that are helper functions either for the solution or some theorem statement. The \texttt{def} is where the problem solution is implemented as a first-order requirement to proving theorems about its correctness.

While a number of theorem provers now exist, we choose Lean because of its explosive growth in mathematics
\footnote{\url{https://leanprover-community.github.io/mathlib\_stats.html}}
and its fertile potential for general-purpose programming~\cite{christiansen2023fpinlean}.
Lean is capable of expressing arbitrary programs,
taking advantage of the \textit{identity monad} and the \texttt{partial} keyword. 

\subsection{Generation Pipeline}

% \begin{figure}
%     \centering
%     \pytestloop
%     \caption{The \texttt{hypothesis}/\texttt{pytest} agent loop. Initialization prompt $p_0$ contains an APPS sample including question, solution, and unit tests alongside a prompt asking the API to generalize the unit tests to \texttt{hypothesis} tests. Model output $c_k$ is run, and if any tests fail all output is formatted and sent back to the model with instructions to address the issues in prompt $p_{k+1}$.}
%     \label{fig:pytest-loop}
% \end{figure}

% \begin{figure}
%     \centering
%     \leanloop
%     \caption{The $\texttt{lean}$ agent loop follows the same structure as the \texttt{pytest} loop, with custom lean prompts and executables.}
%     \label{fig:lean-loop}
% \end{figure}

\begin{figure}
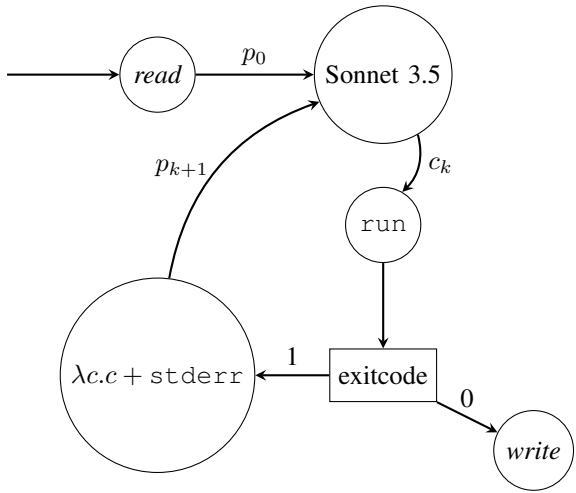

    \centering
    \genericloop
    \caption{Our generic scaffolding loop used at various stages of our pipeline. The \texttt{run} element is replaced with the \texttt{python}, \texttt{pytest}, \texttt{lean}, or \texttt{lake build} executable respectively.}
    \label{fig:generic-loop}
\end{figure}

We use Anthropic's Claude Sonnet 3.5\footnote{Model version \texttt{claude-3-5-sonnet-20241022}, accessed October 25th 2024.} to power five core parts of our benchmark creation pipeline. 
In stages one through three of our pipeline, whenever a non-deductive LLM call is used for processing, we allow for a maximum of 5 attempts before "failing" on that sample. Stage four allows for a maximum of ten, and stage five has no LLM call. These loops include simple instructions that pass relevant error messages back to the model and ask for an updated response.
Given that subsets of our pipeline fall withing the scope of what is capable with current LLMs, we construct scaffolding loops for iterating on model-based solutions.
Broadly, we use outputs from \texttt{python}, \texttt{pytest}, \texttt{lean}, and \texttt{lake build} to ensure the generated code is correct, and to collect standard error and outputs to send back to the model when incorrect (Figure~\ref{fig:generic-loop}).

We first preprocess the original APPS dataset, aggregating unit tests and solutions into one file with a particular function definition and architecture that better lends itself to automated unit testing.
This step also selects a single solution from the set of solutions provided by the original APPS dataset.

Property-based tests are procedurally generated unit tests, up to boundary conditions given on input types~\cite{MacIver2019Hypothesis}. We use our scaffolding loop outlined above, specifically feeding \texttt{pytest}'s standard error and standard out messages back into the next API call. When \texttt{pytest} returns an empty standard error and an exit code of zero, we write the final file of tests to disk and exit the loop.

% Once we have a set of valid and correct Python-based property tests, we enter a similar scaffolding loop (see Figure~\ref{fig:lean-loop}) to construct Lean theorem statements from those tests.

The \texttt{lean} executable is used as our supervisor for the third step, again with standard error and standard output fed back to the model API until a valid
lean file is constructed. The model is prompted to explicitly convert the property tests into unproven theorem statements (cleared via \texttt{sorry}).
The \texttt{lean} executable\footnote{Lean (version 4.12.0, x86\_64-unknown-linux-gnu, commit dc2533473114, Release).} is used to typecheck the theorem statements, ignoring the \texttt{sorry}s but ensuring statements are well-formed. 

Step four of our process aims to add more strict quality assurance to our lean outputs. Using unit tests that have been generated by our preprocessing step as assertions in Python, we deductively parse these assertions to construct inline unit tests in Lean\footnote{\url{https://github.com/leanprover/lean4/blob/master/src/Lean/Elab/GuardMsgs.lean}}.
In parallel, we translate the Python \textit{solutions} into Lean 4 as \texttt{def}s, and require that the guard messages "pass".
While theorem statements and property tests are broadly our primary goal and focus, these unit tests provide an easy way to filter for solutions that are not correct in some minimal form.

The last phase of our pipeline involves ensuring that the theorems we wish to provide are probably true, using Lean's property-based testing framework Plausible\footnote{\url{https://reservoir.lean-lang.org/@leanprover-community/plausible}}. Any theorems that are not "\texttt{plausible}" at this stage, given our solutions that pass unit tests, are filtered.

All code and prompts used for the generation can be found at our open-source GitHub repository.\footnote{\url{https://github.com/quinn-dougherty/fvapps}}

\subsection{Results}

\begin{table}
\centering
\begin{tabular}{|c|c|c|}
    \hline
    $n$ Loops & Pytest $n$ loops & Lean $n$ loops \\
    \hline
    0 & 1963 & 1735 \\
    1 & 1915 & 1450 \\
    2 & 584 & 842 \\
    3 & 143 & 394 \\
    4 & 75 & 183 \\
    5 & 35 & 111 \\
    \hline
\end{tabular}
\vspace{2.5pt}
\caption{Iterations needed for Pytest and Lean agents to converge, filtered to those that made it passed Stage 3 of the pipeline.}
\label{fig:bench-gen-loop-counts} 
\vspace{-10pt}
\end{table}

We provide three subsets of our benchmark according to the assurance level, depending on when the samples were chosen and filtered in our pipeline.

\paragraph{Unguarded} This largest set represents all examples that generated valid theorems, ending at Stage 3 of our full pipeline. Table~\ref{fig:bench-gen-loop-counts} shows the number of execution-generation loops needed to pass our generation criterion in Stages 2 and 3. The Pytest agent took on average 0.846 iterations (with the 0th position as initialization prompt), and the Lean agent took on average 1.188 iterations. 

\begin{figure}
    \centering
    \includegraphics[width=0.9\linewidth]{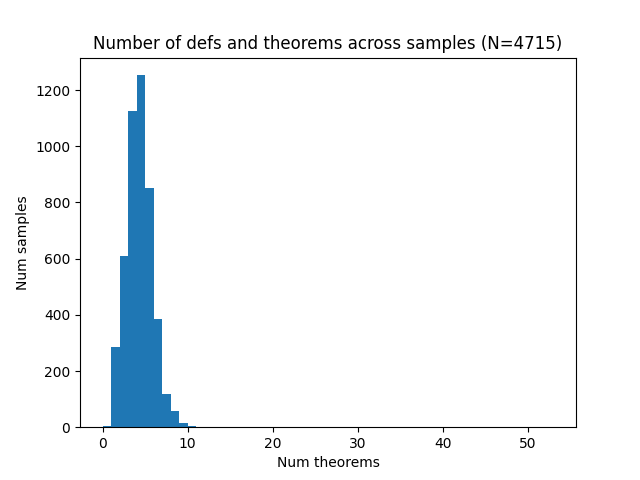}
    \caption{Total number of defs and theorems across FVAPPS samples.}
    \label{fig:benchmark-num-theorems}
\end{figure}

Of the 4715 successful samples, the median number of theorems is four (Figure~\ref{fig:benchmark-num-theorems}).

\paragraph{Guarded} Samples that passed Stage 4 of our pipeline include those samples where solutions were able to be identified that pass simple unit tests.
While we don't include these solutions in our set here, the definitions have been verified to admit a solution such that these basic unit tests pass.\footnote{Importantly this means that \texttt{claude-3-5-sonnet-20241022} was able to solve the problem with our scaffold.}
2089 samples passed this level of assurance.

\paragraph{Guarded and Plausible} 1083 samples passed through all five parts of our pipeline These problems are solvable and well-stated to the highest possible probability short of a full proof (which we leave to the LLMs and other program synthesis methods). Note that the definitions in each of the sets \textit{Guarded} and \textit{Guarded and Plausible} are somehow "easy" in the sense that current language models can provide definitions. 

All 4715 samples are provided as a dataset on HuggingFace with appropriate tagging.\footnote{\url{https://huggingface.co/datasets/quinn-dougherty/fvapps}}

\section{Baseline Experiments}

Our baseline task is to implement and prove each invocation of the word "sorry" in the benchmark's source specification.
% While we aim to provide a benchmark for future models and others in the community to iterate and solve,
We provide preliminary evaluations here, 
using a loop similar to the generation pipeline.
We attempt to solve 101 randomly sampled FVAPPS problems with both Claude 3.5 Sonnet (new) and Gemini 1.5 Pro\footnote{Retrieved in early November 2024.}. For a direct human comparison, we allocated ten hours for a baseliner to attempt one sample.

\subsection{Model Baseline Details}

We use Lean 4.12.0 throughout our benchmark generation and baseline, pinned with Mathlib available in the environment.\footnote{Mathlib commit: 809c3fb3b5c8f5d7dace56e200b426187516535a} We attempt to produce a partial baseline for a subset of FVAPPS using claude-3-5-sonnet-20241022 (the same model that generated the task).

\noindent\textbf{A note on open source models.} We spent some time attempting to solve benchmark samples with small and medium-sized open source models. Particularly, we attempted simple, single-shot autocompletion of theorem statements using DeepSeek's DeepSeek-Prover-V1.5-RL\footnote{https://huggingface.co/deepseek-ai/DeepSeek-Prover-V1.5-RL} \cite{xin2024deepseekproverv15harnessingproofassistant}, but were unable to find a one-shot or loop-based scaffolding solution that made any sufficient progress. An important part of this task seems to be the ability to iterate, and while models fine-tuned to reduce loss on next token prediction in Lean may eventually prove to be useful, currently their limited capability in broader, less narrow goals is insufficient.

\subsubsection{Prompt Details}
Because Lean 4 is in active development and interfaces and functions change with reasonable frequency, we found it was critical to give the LLM sufficient context to avoid common pitfalls associated with biases it may have due to outdated training data. 
We found significantly higher success rates by following a number of key insights by Victor Taelin\footnote{\url{https://github.com/VictorTaelin/AI-scripts/}}, including 1) direct, explicit description of the problem formulation and goal, 2) details of the Lean 4 environment that may be significantly different from its pretrain dataset, and 3) multiple examples of theorem statements and solutions to theorem statements.

\subsubsection{Scaffolding Loop (Agent)}
We initially provide the model with a system prompt consisting of the details in the previous section, and prompt the model first with the original English-language description of the problem and the \texttt{sorry}'d benchmark \texttt{.lean} file (Figure~\ref{fig:sample0023}), with instructions to implement the \texttt{sorry}s. After the model responds, any code block provided is extracted, written to a \texttt{Basic.lean} file, and run in a custom managed environment with specifically pinned libraries using \texttt{lake build}. The full standard input and output alongside returncode are sent back to the model with the instruction to address the issues.

Initial attempts were unsuccessful in solving all of the theorem statements at once; to give the model a better chance at solving the problem we show and allow the model to solve one statement at a time, starting with the definitions and then appending each theorem independently, to be proven separately from the rest.

\subsection{Human Baseline}

A human attempted the function and one of the theorems from sample 0023. It took them 10 hours to accomplish the function shown in Figure \ref{fig:humimpl0023}, since they needed termination proofs for two recursors. In the time allotted, they did not make it through any of the proofs. A problem they ran into with the proofs was stack overflow trying to unfold the recursors. Given this, we don't think a full human baseline is a feasible option.

\begin{figure}
   \centering
\begin{lstlisting}
def go (voters holdouts : List (Nat × Nat)) (expense conversions : Nat) : Nat :=
  if conversions ≥ voters.length then
    expense
  else
    let filtered := filter_cascade holdouts conversions
    if filtered.isEmpty then
      expense
    else
      let bestValue := filtered.argmax? utility |>.get!
      let holdouts' := filtered.eraseP (. == bestValue)
      go voters
         holdouts'
         (expense + bestValue.snd)
         (conversions + 1)
  termination_by (voters.length - conversions)
  decreasing_by omega

def solve_elections (n : Nat) (voters : List (Nat × Nat)) : Nat :=
  go voters voters 0 0
\end{lstlisting}
   \caption{Human baseline's solution to the definition for sample 23 of FVAPPS (helper recursor omitted for brevity, available on GitHub)}
   \label{fig:humimpl0023}
\end{figure}

\subsection{Comparing Model Baseline and Human Baseline on One Sample}

On sample 23 regarding the crooked election, both successfully implemented the definition of the function (Appendix \ref{appndx:solutions-23}) and succeeded on no nontrivial proofs. We let Sonnet attempt the second theorem for 100 loops, and while it made a valiant effort, it diverged. The human and Sonnet clearly took two very different approaches, but it remains to be seen which approach has better proof ergonomics.  

\subsection{Model performance on more samples}
\begin{figure}
    \centering
    \includegraphics[width=0.9\linewidth]{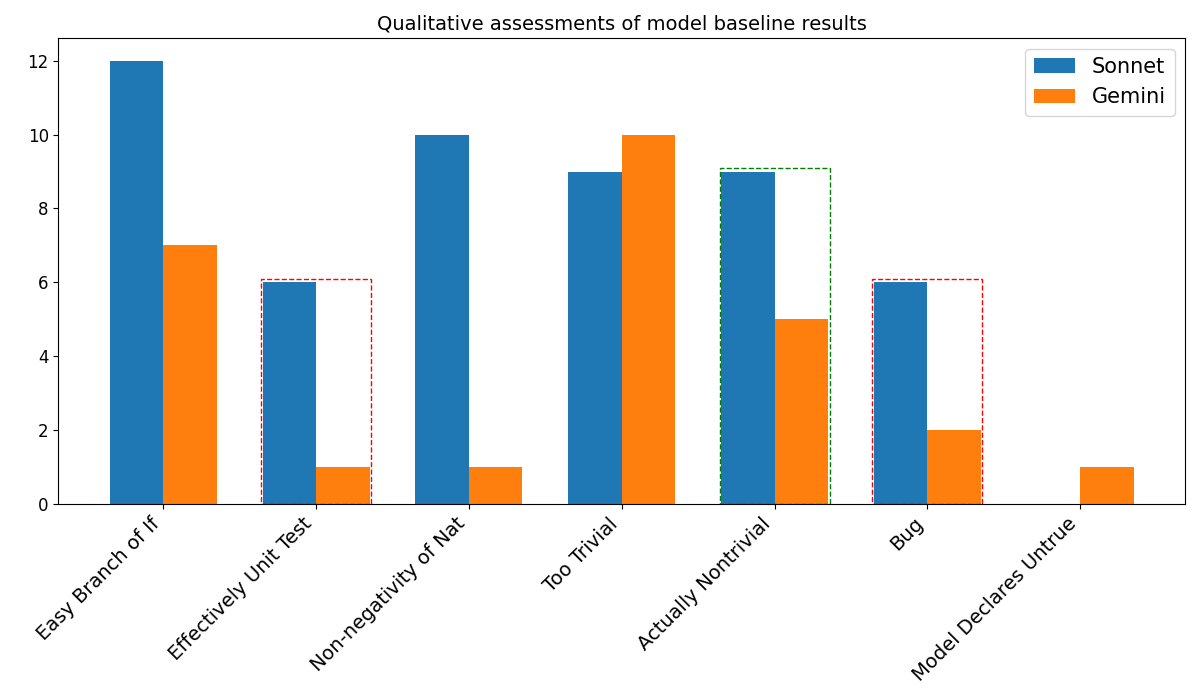}
    \caption{Qualitative categories of theorem solutions in the 100 samples, first two theorems each sample. The red box shows completely spurious results, either bugs or a substitution of a quantified variable with a single value. The green box shows the most nontrivial results. The other categories are neither spurious nor impressive, though they require some syntactic fluency that many language models would fail at.}
    \label{fig:qualitative-bar-chart}
\end{figure}

Both Sonnet and Gemini ran for a maximum of 25 loops for the \texttt{def}s and a max of 50 loops for each \texttt{theorem}.
Sonnet completed all but seven of the definitions and 43 samples had at least one theorem completed. Gemini completed only 71 of the definitions, and 23 samples had at least one theorem completed. Figure \ref{fig:qualitative-bar-chart} shows the breakdown across our different buckets, assessed qualitatively. The green box surrounds the best solutions, nontrivial proofs that involve both intricate reasoning and correct syntax. The red box shows completely spurious results, either bugs or a substitution of a quantified variable with a single value. The other categories are neither spurious nor impressive, though they require some syntactic fluency that many language models would fail at. Importantly, Gemini declared the second theorem of sample 3078 incorrect/unsolvable in a comment, while Sonnet nontrivially solved it. 

\begin{figure}
    \centering
    \includegraphics[width=0.9\linewidth]{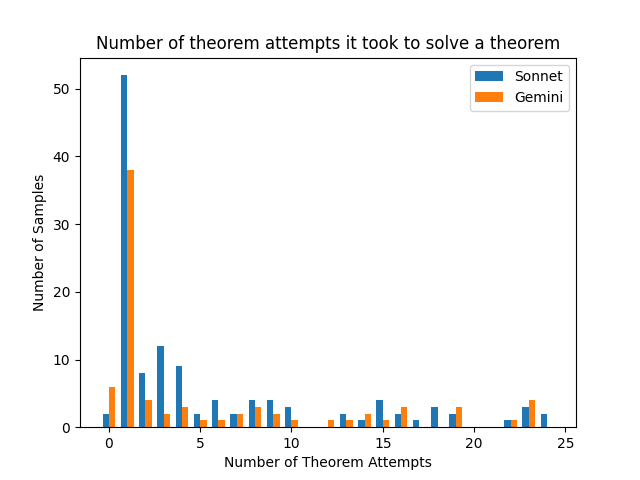}
    \caption{Number of theorem attempts it took to solve a theorem, conditional on that theorem succeeding.}
    \label{fig:theorems-attempts}
\end{figure}

Figures \ref{fig:theorems-attempts} and \ref{fig:defs-attempts} show how many samples took how many attempts. Of the theorems that got eventually completed, roughly 20\% of each model's were done in zero or one iteration of the loop. 

Of the 406 theorems attempted, Sonnet accomplished 121 ($30\%$) and Gemini accomplished 74 ($18.5\%$).

\subsection{Discussion}

We noticed that Sonnet solved many problems via imperative programming with the identity monad (\texttt{Id.run do}). However, this kind of solution fails to capture inductive structure that lends itself to proofs. This means future solutions to FVAPPS will take one of two directions: an unwieldy and uninterpretable proof involving the identity monad, or ensuring that solutions are factored in a way that can exploit inductive structure. 

Table~\ref{tab:baseline-splits} shows the split over our quality assurance steps over the entire baseline set and those that were solved. While this sample is small, we roughly find that our higher-quality samples that are both guarded and plausible are somewhat more likely to be solvable by current models.

\begin{table}[]
    \centering
    \begin{tabular}{c|c|c|c}
        \hline
        \hline
         & Baseline & Sonnet    & Gemini  \\
         & Counts   & Successes & Successes  \\
         \hline
        Unguarded & 69 & 41 & 28 \\
        Guarded & 18 & 12 & 11 \\
        Guarded and Plausible & 14 & 7 & 4 \\
        \hline
        Total & 101 & 60 & 43 \\
        \hline
        \hline
    \end{tabular}
    \caption{Baseline results split across quality assurance stages of our generation pipeline.}
    \label{tab:baseline-splits}
\end{table}

\section{Related Work}

Research and development into stronger programming guarantees spans a number of large and growing communities in the software landscape.
While interest in generally stronger-typed programming paradigms is growing (e.g., growing interest in converting the Linux kernel codebase to Rust,\footnote{\url{https://github.com/Rust-for-Linux/linux}}
increasing availability of static typing in Python\footnote{\url{https://realpython.com/python312-typing/}}) there remains a smaller group of people and programming languages focused directly on formal verification.
Coq~\cite{coq}, Dafny~\cite{dafnybook}, Isabelle~\cite{isabellebook}, and Lean~\cite{demoura2015lean} all target theorem proving and formal verification as first class goals, albeit with varying visions and core foundations.

Automating programming with machine learning or program synthesis is a rich area~\cite{ellis2020dreamcodergrowinggeneralizableinterpretable, austin2021programsynthesislargelanguage}

\subsection{Datasets and Benchmarks}

Informal Mathematical theorem proving datasets are the most extensive, with benchmarks ranging from 15,000 to over 100,000 proofs~\cite{coqgym,leandojo,jiang2021lisa,naturalproofs}, plus the Archive of Formal Proofs containing around 1 million lines of code~\cite{afp}. Unverified program synthesis benchmarks are typically an order of magnitude smaller, containing thousands of programs, with APPS being the largest at 10,000 programs~\cite{Hendrycks2021MeasuringCC} (though LiveCodeBench continues to grow~\cite{livecodebench}). In stark contrast, formal software verification benchmarks have historically been much more limited in size, with both available datasets (Clover and dafny-synthesis) containing order of 100 programs each~\cite{clover,dafny-synthesis}, with DafnyBench~\cite{loughridge2024dafnybench} up at 782.

\subsection{Models}

Models like DeepSeek \cite{xin2024deepseekproveradvancingtheoremproving,xin2024deepseekproverv15harnessingproofassistant} and AlphaProof \cite{deepmind2024imo} have advanced the ability of AI to do math, both formally and informally. DeepSeek Prover has demonstrated significant capabilities in formal mathematics, achieving state-of-the-art performance on the Lean theorem proving benchmark. The model combines large-scale pretraining on mathematical texts with specialized fine-tuning on formal proof corpora, enabling it to generate valid formal proofs for complex mathematical statements.

AlphaProof~\cite{deepmind2024imo} represents another milestone in automated theorem proving, showing particular strength in creative mathematical problem-solving. Its success on IMO-level problems demonstrates the potential for AI systems to engage in sophisticated mathematical reasoning, though the gap between informal mathematical proofs and formal verification remains significant.

In the domain of program synthesis and verification, models have evolved along several tracks. Traditional program synthesis approaches using symbolic techniques and SMT solvers have been augmented by neural methods~\cite{wang2023legoprover}. Large language models fine-tuned on code, such as CodeLlama \cite{Rozire2023CodeLO} and GPT-4\cite{gpt4}, have shown promising results in generating functionally correct programs, but their capabilities in formal verification remain limited.

Recent work has begun to bridge this gap \cite{Baif2024DafnyCopilot, yang2024autoverusautomatedproofgeneration} demonstrating the potential for AI assistance in formal program verification. However, these systems typically require significant human guidance and cannot yet autonomously generate both correct programs and their formal proofs. The challenge of simultaneously reasoning about program semantics and verification conditions remains an open problem in the field.

\section{Discussion}

Our current presentation has a few limitations, and clear directions for future work.
While the samples in our final \textit{guarded} and \textit{plausible} set are quite high quality, it is nontheless possible that the theorem statements do not correspond to desirable properties nor that they, in union, cover the full range of properties one would ideally prefer.
While manual spot-checking did not reveal any systematic failures post complete five-stage processing,
it is nontheless possible that particular theorems are not applicable or vacuous in some form.

The FVAPPS benchmark evaluates AI systems on both program synthesis and formal verification by extending APPS with Lean 4 theorem specifications. While current language models can implement programming solutions\footnote{https://www.anthropic.com/news/3-5-models-and-computer-use}\footnote{https://openai.com/index/introducing-swe-bench-verified/}, they struggle with formal verification proofs, highlighting the challenge between imperative programming and proof-amenable structures. The benchmark's difficulty is demonstrated by limited progress from both human baselines and models. FVAPPS opens research directions in simultaneous implementation and verification reasoning, automated proof-amenable solution structuring, and scaling formal verification, providing a framework to measure progress toward verified automated programming.

\section*{Acknowledgment}

We thank Buck Shlegeris, Jason Gross, and Rajashree Agarwal for valuable feedback and discussion during early versions of this work.
We are also grateful to the Anthropic External Researcher program for providing API credits for benchmark generation, and to FAR.ai for providing operations and administrative support during the course of the project.

\appendix

\subsection{Details of buckets in \ref{fig:qualitative-bar-chart}}

Table \ref{tab:model-comparison} declares which samples in particular fall into Figure \ref{fig:qualitative-bar-chart}'s buckets. Of note is that 3078 is declared untrue in a code comment by Gemini, but solved completely by Sonnet. Otherwise, many samples fall into the same buckets across models. 

\begin{table*}[ht]
\centering
\begin{tabular}{|c|p{5cm}|p{5cm}|}
\hline
Category & Sonnet & Gemini \\ \hline
Easy branch of if & 0282, 3430, 4430, 4172, 3057(0), 3040, 0375(0), 1265, 3430, 2052, 0267(0), 1318(1) & 4430, 3040, 0375, 3199, 3430, 0267(0), 1318 \\ \hline
Effectively unit test & 1804, 2459, 1233, 0666, 1732, 0267(1) & 0267 \\ \hline
Non-negativity of Nat & 2459, 1233, 4742, 0375(1), 1850, 4036, 0077, 0069, 0669, 1318(0) & 0077 \\ \hline
Too trivial & 3819, 3192, 2667, 1306(0), 1306(1), 2649, 2109, 4002, 4350 & 3151, 3819, 4017, 0986(0), 0986(1), 4045, 2649, 2112, 2109, 4350 \\ \hline
Actually nontrivial & 3057(1), 3402, 0314, 1710, 4412, 0323, 4045, 2342, 3078 & 3057, 3402, 1947(0), 3078(0), 2052 \\ \hline
Bug & 2459, 4017, 0930, 4802, 3199(0), 3199(1) & 1947(1), 4466 \\ \hline
Model declares untrue & & 3078(1) \\ \hline
\end{tabular}
\vspace{2.5pt}
\caption{Comparison of theorem classifications between Sonnet and Gemini}
\label{tab:model-comparison}
\end{table*}

\subsection{Number of definition attempts it took to solve a function, conditional on it succeeding}

\begin{figure}[h]
    \centering
    \includegraphics[width=0.9\linewidth]{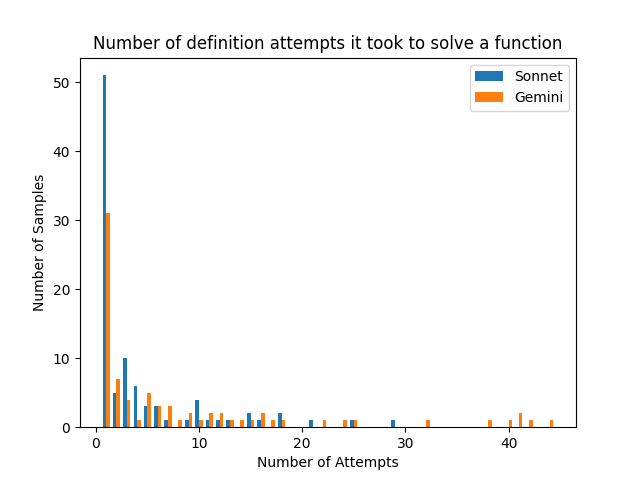}
    \caption{Definition attempts needed to solve a function, conditional on success.}
    \label{fig:defs-attempts}
\end{figure}

Figure \ref{fig:defs-attempts} is a companion to Figure \ref{fig:theorems-attempts} but refers to definitions instead of theorems. Of the \texttt{def}s that succeeded at all, 50\% of Sonnet's and 31\% of Gemini's succeeded in one shot. 

\subsection{Baseline solutions to problem 23}\label{appndx:solutions-23}

Gemini accomplished \ref{fig:geminiimpl0023} in 11 attempts and sonnet failed \ref{fig:sonnetimpl0023} after 50 attempts.

\begin{figure}
    \centering
\begin{lstlisting}
def solve_elections (n : Nat) (voters : List (Nat × Nat)) : Nat := Id.run do
  let range := List.range (n + 1)
  let mut best := none
  for k in range do
    match try_subset ⟨n, voters, 0, k, []⟩ with
    | some cost => 
      match best with
      | none => best := some cost
      | some b => if cost < b then best := some cost
    | none => continue
  best.getD 0
\end{lstlisting}
    \caption{claude-3-5-sonnet-20241022's attempt at the definition for sample 23 of FVAPPS (\texttt{try\_subset} omitted for brevity, available on GitHub)}
    \label{fig:sonnetimpl0023}
\end{figure}

\begin{figure*}
    \centering
\begin{lstlisting}
def solve_elections (n : Nat) (voters : List (Nat × Nat)) : Nat :=
  let voters := voters.toArray.qsort (fun v₁ v₂ => (v₁.fst < v₂.fst) || (v₁.fst == v₂.fst && v₁.snd > v₂.snd)) |>.toList  -- Use boolean comparisons for sorting
  let rec go (cost : Nat) (numVoters : Nat) (remainingVoters : List (Nat × Nat)) : Nat :=
    if numVoters == n then cost
    else
      match remainingVoters with
      | [] => cost
      | (m, p)::rest =>
        let buyCost := go (cost + p) (numVoters) rest
        let freeCost := if m <= numVoters then go cost (min n (numVoters + 1)) rest else buyCost
        min buyCost freeCost
  go 0 0 voters
\end{lstlisting}
    \caption{gemini-1.5-pro's solution to the definition for sample 23 of FVAPPS}
    \label{fig:geminiimpl0023}
\end{figure*}

%\clearpage
%\newpage
\printbibliography

\end{document}